\documentclass[useAMS,usenatbib]{mn2e}

\usepackage{graphicx}
\usepackage[fleqn]{amsmath}

\newcommand{\msun}{\mathrm{M_\odot}}
\newcommand{\ab}{a_\mathrm{b}}
\newcommand{\eb}{e_\mathrm{b}}

\newcommand{\ef}{e_\mathrm{f}}
\newcommand{\vesc}{v_\mathrm{esc}}
\newcommand{\gamceph}{\gamma~\mathrm{Cephei}}
\newcommand{\alphacen}{\alpha~\mathrm{Centauri}}
\newcommand{\rinf}{R_\mathrm{infl}}
\newcommand{\mmin}{M_\mathrm{min}}
\newcommand{\sigd}{\Sigma_\mathrm{d}}
\newcommand{\tauctd}{\tau_\mathrm{c,2D}}
\newcommand{\tauc}{\tau_\mathrm{c}}

\title{Planetesimal collisions in binary systems}

\author[S.-J. Paardekooper and Z.M. Leinhardt]{S.-J. Paardekooper$$\thanks{E-mail: 
S.Paardekooper@damtp.cam.ac.uk} and Z.M. Leinhardt\\  
DAMTP, University of Cambridge, Wilberforce Road, Cambridge CB3 0WA,
United Kingdom}

\begin{document}

\date{Draft version \today}

\pagerange{\pageref{firstpage}--\pageref{lastpage}} \pubyear{2009}

\maketitle

\label{firstpage}

\begin{abstract}
We study the collisional evolution of km-sized planetesimals in tight binary star systems to investigate whether accretion towards protoplanets can proceed despite the strong gravitational perturbations from the secondary star. The orbits of planetesimals are numerically integrated in two dimensions under the influence of the two stars and gas drag. The masses and orbits of the planetesimals are allowed to evolve due to collisions with other planetesimals and accretion of collisional debris. In addition, the mass in debris can evolve due to planetesimal-planetesimal collisions and the creation of new planetesimals. We show that it is possible in principle for km-sized planetesimals to grow by two orders of magnitude in size if the efficiency of planetesimal formation is relatively low. We discuss the limitations of our two-dimensional approach.
\end{abstract}
 


\section{Introduction}
We consider the problem of planetesimal evolution in tight binary star systems such as $\gamceph$ and $\alphacen$. To date several hundred extrasolar planets have been detected -- 20$\%$ of which orbit the primary of a binary or multiple system \citep{desi07}. In most of these cases the stars are widely separated and do not significantly affect the evolution of planets, but $\gamceph$, Gl86, HD41004A, and possibly HD196885 are examples of tight binary systems with a semi-major axis, $\ab$, around 20 AU, that have Jupiter mass planets. $\gamceph$ has the most extreme binary parameters of the known tight binaries with detected planets. The primary star is a K giant ($M_\mathrm{A \star} \sim 1.4$ $\msun$) the secondary is a M dwarf ($M_\mathrm{B \star} \sim 0.4$ $\msun$), the binary has a high eccentricity ($\eb=0.4$), and a small $\ab$ of 20 AU. The planet is 1.6 $\mathrm{M_J}$ with an $a \sim 2$ AU \citep{neu07}. Due to its extreme characteristics $\gamceph$ is often considered one of the most stringent tests for planet formation models. 

Previous work by \citet{the04} and \citet{Quintana2007} has shown that
planets can form in $\gamceph$ if planetary embryos can form within
the stable region $< 3$ AU \citep{Wiegert1997}. However, work on the
earlier stage of planet formation has shown that it is very difficult to grow planetesimals that are initially in the km size regime because of the secular perturbations from the secondary star on the planetesimals and the gas disc. The coupled effect of perturbations from the binary and the gas drag from the disc causes differential orbital phasing, and high speed destructive impacts between planetesimals of different sizes. \citet{thebault06} suggest that the destructive collisions prohibit the growth of planetesimals in $\gamceph$ at 2 AU, the current location of the planet.

Although planets have yet to be detected in the $\alphacen$ system, the proximity of the system and the roughly solar mass of both stellar components make the system a prime target for finding an Earth mass planet in the habitable zone \citep{Guedes2008}. However, the $\alphacen$ system is more extreme than $\gamceph$ with a mass ratio close to one ($M_{A \star} = 1.1$ $\msun$ and $M_{B \star} = 0.93$ $\msun$), $\eb = 0.52$, and $\ab = 23.4$ AU. \citet{Thebault2008,Thebault2009} found that impact speeds between different sized planetesimals would be above the disruption threshold due to differential orbital phasing, and planetesimals could not grow in the habitable zone in either $\alphacen$ A or B. 

In this letter we show that planetesimal growth from 1 to 100 km is possible in principle in these tight binaries when the collisional evolution of the planetesimal disc is taken into account. Disruptive collisions will prevent differential orbital phasing from being set up and, at the same time, create a reservoir of small debris that can be accreted onto the left over planetesimals. 

\section{Differential orbital phasing}

\begin{figure} 
\resizebox{\hsize}{!}{\includegraphics{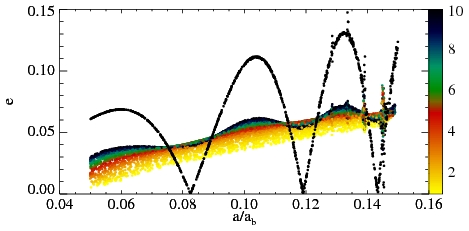}} 
\caption{Eccentricity vs. semi-major axis of collisionless, massless test particles in the $\gamceph$ system after 100 binary orbits in a gas disc with a constant gas density of $\rho_\mathrm{g}=1.4\cdot 10^{-9}$ g $\mathrm{cm^{-3}}$. Colour indicates size in km; the black curve with the large oscillations denotes a gas-free case.} 
\label{figdiffphase} 
\end{figure} 

In the absence of any gas, a tight eccentric binary companion will excite eccentricities of particles orbiting the primary. For particles starting on circular orbits, their eccentricity will oscillate around the forced eccentricity $\ef$:
\begin{equation}
\ef(a)=\frac{5}{4}\frac{a}{\ab}\frac{\eb}{1-\eb^2},
\end{equation}
with amplitude $2\ef$, where $a$ is the semi-major axis of the particle \citep{heppenheimer78}. Since there is no dissipative force in the system, the oscillations do not damp and their spatial frequency increases with time, which eventually leads to orbital crossing of neighbouring particles \citep{thebault06}. This is illustrated by the black curve in Fig. \ref{figdiffphase}. As soon as particles at an eccentricity maximum can collide with particles at an eccentricity minimum the resulting encounter velocities are too high for accretion to occur.

The presence of a gas disc and the associated aerodynamic gas drag on the particles can effectively damp the oscillations leading to a well-defined equilibrium eccentricity distribution $e_0=e_0(a)$ \citep[see][]{paard08}, with corresponding periastron alignment. The time scale for this equilibrium to be reached is the secular time scale. Therefore, it takes many binary orbits before planetesimals are on their equilibrium orbits. Once they have reached their equilibrium orbits particles of the same size have orbits that are in phase and, therefore, they will not suffer from high-speed collisions with each other \citep{Marzari2000}. However, since the magnitude of the gas drag force depends on the particle size, each particle size will have a different $e_0(a)$, a phenomenon that is called differential orbital phasing \citep{thebault06}. In Fig. \ref{figdiffphase} particles of a single size (and colour) follow phased orbits with no oscillations. The residual oscillations that can be seen for the largest particles will eventually be damped completely by gas drag. However, for particles of differing size there is a considerable spread in eccentricity and, not shown, longitude of periastron. The difference in eccentricity magnitude can easily exceed $0.01$ for particles of 1 and 5 km. This implies that particles of different size will undergo high-speed collisions, which are usually destructive. As a result, it is difficult to form planetary cores in tight binary systems if the only growth mechanism is accretionary planetesimal-planetesimal collisions.

\section{Collision time scale}

\subsection{Single star case}
Consider a population of equal-sized bodies of mass $M$ around a single star at 1 AU. The collision time scale is given by $\tauc=1/(\pi R^2 n \Delta v),$
where $R$ is the radius of the bodies, $n$ the number density and $\Delta v$ the velocity dispersion. The latter will be of the order of the escape velocity, $\vesc=\sqrt{2\mathrm{G}M/R}$. Focusing on $a=1$ AU around a Solar type star we have $v_\mathrm{esc}=4.5\cdot 10^{-5} (R/\mathrm{km}) a\Omega$, where $\Omega$ is the angular velocity. The number density $n$ is given by $n=\Sigma_\mathrm{s}/(\Delta z M)$, where $\Sigma_\mathrm{s}$ is the surface density of solids and $\Delta z$ is the vertical extent of the particle disc. We can write $\Delta z\approx a i$, with $i$ the maximum inclination, which we can assume to be small. In fact, we expect $2i\approx e \approx \vesc/a\Omega$. Plugging this all in we find for the collision time scale:
\begin{equation}
\tauc\approx11750 \frac{R}{\mathrm{km}} \frac{17~\mathrm{g~cm^{-2}}}{\Sigma_\mathrm{s}}\frac{\rho_\mathrm{p}}{3~\mathrm{g~cm^{-3}}}\Omega^{-1},
\end{equation}
where $\rho_\mathrm{p}$ is the bulk density of the particles. This means that for typical parameters at 1 AU, a km-sized planetesimal will undergo a collision once every few thousand years.

\subsection{Binary star case}
Perturbations due to a coplanar, eccentric binary will increase the eccentricity of the planetesimals. Differential orbital phasing can easily give rise to the velocity dispersion entering the collision time scale will go up by two orders of magnitude compared to the single star case. Keeping all other parameters the same, a km-sized planetesimal will now undergo a collision every 10-20 orbits at 1 AU. If we take the binary companion to have a semi-major axis of 20 AU, this collision time scale amounts to $25~\%$ of a binary orbit. This should be compared to the time scale for a particle to reach its equilibrium orbit in the binary system, which happens on a secular time scale (many binary orbits). Therefore, physical collisions are of crucial importance for the evolution of the planetesimal population. It is important to stress that not only, as has been realised before, are collisions destructive when differential orbital phasing happens, they can actually prevent this size-dependent orbital structure from being set-up in the first place. It is therefore necessary to take collisions into account when studying the effect of differential orbital phasing on planetesimal accretion. 

Another consequence of having many possibly catastrophic collisions in the system is a large amount of small debris. This small debris can act as a source of new planetesimals or be accreted onto remaining larger bodies. All these ingredients need to be taken into account in a consistent model of planetesimal accretion in binary systems.

\subsection{Two-dimensional disc}
In order to reduce the computational cost of the models, we will work in a two-dimensional (2D) geometry. The collision time scale, for orbits with random phase, is then given by
\begin{equation}
\tauctd=\frac{1}{2R\bar n \Delta v}=\frac{2.4\cdot 10^{-4}}{e}\left(\frac{R}{\mathrm{km}}\right)^2 \frac{17~\mathrm{g~cm^{-2}}}{\Sigma_\mathrm{s}}\frac{\rho_\mathrm{p}}{3~\mathrm{g~cm^{-3}}}\Omega^{-1},
\end{equation}
where $\bar n$ is the surface number density and $e$ is a measure of the velocity dispersion ($e\approx 4.5\cdot 10^{-5}$ if $\Delta v=v_\mathrm{esc}$ for 1 km sized bodies). The collision time scale for a 2D disc is artificially short; we have to correct for this by choosing a low surface density in the simulations so that for the unperturbed disc, $\tauctd=\tauc$. 

\section{Model design}
From the above discussion it is clear that the effect of collisions, on the size distribution as well as on the orbital elements of the planetesimals, can not be ignored. As a result, we use a simple model that incorporates the three necessary components: gas, planetesimals and small dust.

\subsection{Gas disc}
For simplicity, we take the gas disc to be static and circular. Although the gas disc is expected to become eccentric under the influence of the binary companion \citep{paard08,kley08} the qualitative outcome of the model presented here is not affected by the assumption of zero eccentricity.  \cite{paard08} showed that if the gas disc does not follow the forced eccentricity, differential orbital phasing will occur, which always occured in the full hydrodynamical simulations. A circular gas disc is, therefore, a reasonable starting point.

The gas density used in this work is assumed to be constant with $a$. The value is chosen to be close to the Minimum Mass Solar Nebula at 1 AU, $\rho_1=1.4\cdot 10^{-9}~\mathrm{g~cm^{-3}}$. The resulting gas drag force is
\begin{equation}
{\vec F}_\mathrm{drag}=-\frac{3\rho_\mathrm{g} C_\mathrm{d}}{8\rho_\mathrm{p}R}
|{\vec v}-{\vec v}_\mathrm{g}|({\vec v}-{\vec v}_\mathrm{g}),
\end{equation}
with $C_\mathrm{d}$ the drag coefficient. We take $C_\mathrm{d}=0.4$, appropriate for spherical bodies.

\subsection{Small dust}
We embed the gas disc with a population of tightly coupled small particles, which are assumed to move on circular orbits together with the gas. They can be accreted onto existing planetesimals, but also form new planetesimals, and are created in destructive collisions between planetesimals. The small dust particles are distributed over typically 32 radial bins. At the start of the simulation, the mass in dust is equal to $M_\mathrm{d}=\mmin N$, where $\mmin$ is the mass of a planetesimal of the minimum size we consider (usually 1 km), and $N$ is an input parameter specifying the total mass. Note that $N$ is also the maximum number of particles we could have at any one time in the simulation. The dust mass inside a radial bin is smeared out to give a smooth surface density $\Sigma_\mathrm{d}$.
 
\subsection{Planetesimals}
We model planetesimals as test particles, moving under the influence of gravity from both stars as well as gas drag. Planetesimals have a minimum size of 1 km (everything smaller is taken to be small dust), and can grow by low-velocity collisions with each other and by accreting small dust. Collisions are detected by assigning an inflated radius $\rinf \propto R$ to the particles and testing whether two particles overlap.  The planetesimal disc is characterised by the product $\rinf N$, which, together with $\Delta v$, sets the collision time scale. Since $\Delta v$ is set by the gravitational perturbations due to the binary companion and is independent of $N$ and $\rinf$, we expect simulations with the same value for  $\rinf N$ to give similar results, a well-known result of collision models.

\subsection{Collision outcomes}
We use the velocity-dependent catastrophic disruption criterium of \cite{stewart09} (strong particle version) to determine the size of the largest remnant when a collision is detected. We then use the technique described in \cite{wyatt02} to predict the second-largest remnant. The total mass minus the largest remnant is assumed to follow a power law size distribution with index $-1.93$, from which the number of bodies larger than $R$, $N(>R)$, can be derived. The size for which $N(>R)=2$ is the size we take for the second largest remnant \citep{wyatt02}. Any mass that is below the minimum particle size is added to the small dust. To keep the total number of particles tractable, we do not keep the third largest remnant, even if it is bigger than the minimum size.

\subsection{Planetesimal formation}
Contrary to all previous studies, we do not let all planetesimals appear at $t=0$, but instead have them form from the small dust present in the system. The efficiency of planetesimal formation is a major unknown in this model. We take a very simple approach and say that:
\begin{equation}
\frac{dM_\mathrm{A}}{dt}=2\epsilon_\mathrm{p}R_\mathrm{infl,min}\frac{M_\mathrm{d}}{\mmin}\Sigma_\mathrm{d}a\Omega,
\end{equation}
where $\epsilon_\mathrm{p}$ is an efficiency factor, $R_\mathrm{infl,min}$ is the inflated radius of the minimum planetesimal size, and $M_\mathrm{A}$ is the mass available (in a certain radial dust bin) to form planetesimals. As soon as $M_\mathrm{A} > \mmin$, we create a new planetesimal. Note that for fixed $\rinf N$, $M_\mathrm{A} \propto N$, so that the number of planetesimals at any time is proportional to $N$. Increasing $N$ while keeping $\rinf N$ constant effectively increases the 'resolution' of the simulation. 

For equal masses in km-sized planetesimals and small dust, the collision time scale and the planetesimal formation time scale are related through $\tau_\mathrm{p}=\tauc e/\epsilon_\mathrm{p}$. If we consider the single star case, with $e\approx 10^{-5}$ and $\tauc\approx 10^3$ yr, and if we expect planetesimal formation to proceed on a time scale of $\tau_\mathrm{p}>10^4$ yr, we must have $\epsilon_\mathrm{p} \sim 10^{-6}$. This is the value used for the standard model discussed below.

\subsection{Dust accretion}
Planetesimals can accrete dust at a rate
\begin{equation}
\frac{dM}{dt}=2\epsilon_\mathrm{d}\rinf\sigd |{\vec v}-{\vec v}_\mathrm{g}|,
\end{equation}
where $\Sigma_\mathrm{d}$ is the surface density of small dust. An efficiency factor $\epsilon_\mathrm{d}$ accounts for the possibility that not all dust created in collisions is available for accretion, and that the efficiency of accreting small bodies may be smaller than 1. Note that $\sigd \propto N$, so that the amount of dust accreted per planetesimal is the same if $N$ is changed but the product $\rinf N$ is kept constant. 

\section{Results}

\begin{figure} 
\centering
\resizebox{\hsize}{!}{\includegraphics[]{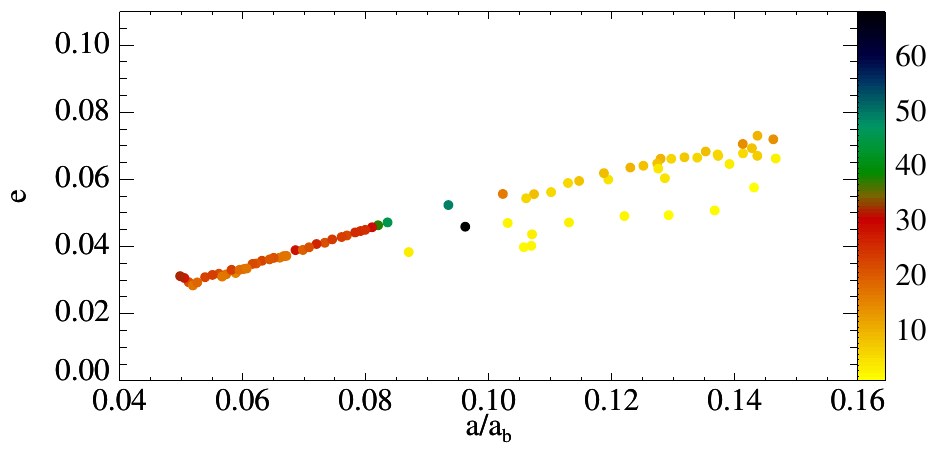}} 
\caption{Eccentricity vs. semi-major axis of test particles in the $\gamceph$ system after 1000 binary orbits. Color indicates size in km.} 
\label{figae} 
\end{figure} 

We start by describing a simulation that has $N=10^6$, $\rinf/\ab=10^{-5}R/\mathrm{km}$, $\epsilon_\mathrm{p}=10^{-6}$, $\epsilon_\mathrm{d}=1$ and constant $\rho_\mathrm{g}=1.4\cdot 10^{-9}$ g $\mathrm{cm^{-3}}$. Note that gas drag, and therefore differential orbital phasing, is very strong throughout the disc. The minimum particle size is 1 km and there are no planetesimals at $t=0$. Binary parameters are those of the $\gamceph$ system. Figure \ref{figae} shows the resulting $(a,e)$ distribution after 1000 binary orbits ($\sim 90000$ yr). It is clear that significant accretion has taken place in the inner parts of the disc, where particles have grown from 1 km to 50 km. Outside $a/\ab=0.1$, which corresponds to $2$ AU, perturbations due to the binary are too strong for accretion to occur. 

\begin{figure} 
\resizebox{\hsize}{!}{\includegraphics{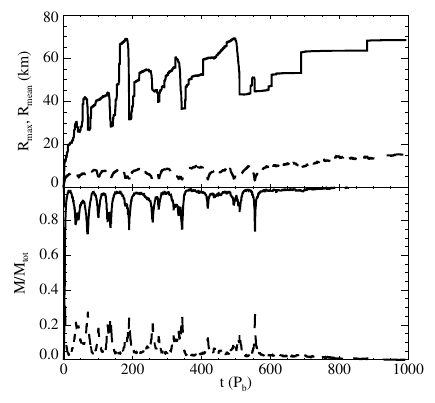}} 
\caption{Evolution of maximum particle size (solid curve, top panel), mean particle size (dashed curve, top panel), planetesimal mass fraction (solid curve, bottom panel) and dust mass fraction (dashed curve, bottom panel) for the $\gamceph$ system.} 
\label{figsizemass} 
\end{figure} 

Inside $2$ AU, planetesimals grow to sizes up to $70$ km. The top panel of Fig. \ref{figsizemass} shows the time evolution of the maximum and mean planetesimal size over the whole disc. The mean size is dominated by the large number of small planetesimals in the outer disc, and stays between $10$ and $20$ km. The maximum particle size goes up rapidly by dust accretion (smooth parts of the curve) and by accreting collisions (jumps). The largest planetesimal is destroyed a few times as well, but in general the trend is to grow to larger sizes. After $\sim 600$ binary orbits, there is no source of small dust remaining from which to create new planetesimals, and the collision time scale goes up. 

The maximum size that can be reached in this scenario is limited by the total amount of solid material present in the disc. In the 2D approximation, this mass has to be artificially low in order to end up with a realistic collision time scale. For a disc that has twice the solid material, keeping all other parameters the same, growth up to 100 km was observed. Increasing $N$ by a factor of 2 while decreasing $\rinf$ by the same factor, which amounts to increasing the resolution of the simulation, showed growth up to 150 km. In these higher resolution runs more accreting collisions are observed than depicted in Fig. \ref{figsizemass}. This is because the system goes through phases of low particle number density, for which the collision statistics in the lower resolution runs are not optimal. In this sense, Fig. \ref{figsizemass} represents a worst-case scenario, where many low-velocity collisions are missed, and adding particles, while keeping the collision time scale the same, will only favour planetesimal growth more.

Gas drag appears to play only a minor role in determining the qualitative outcome of the model. This is mainly due to the fact that the planetesimals are weak enough so that any small eccentricity difference of $\sim 0.01$ will lead to destructive collisions. Whether this difference is due to differential orbital phasing (under the influence of gas drag) or simply orbital crossing (in the absence of gas) does not matter. A simulation without any gas showed the same trend as in Fig. \ref{figsizemass}, with growth up to 80 km. 

Crucial parameters are the efficiency of planetesimal formation and dust accretion. Accretion as shown in Fig. \ref{figsizemass} can only occur if the small debris created in catastrophic collisions is swept up by larger bodies rather than forming new planetesimals. Increasing $\epsilon_\mathrm{p}$ by a factor of 10 still results in accretion up to 80 km, but for a factor 100 growth stalls at 10 km. Similarly, the result depicted in Fig. \ref{figsizemass} is robust to changes in $\epsilon_\mathrm{d}$ up to a factor of 10. 

\begin{figure} 
\resizebox{\hsize}{!}{\includegraphics{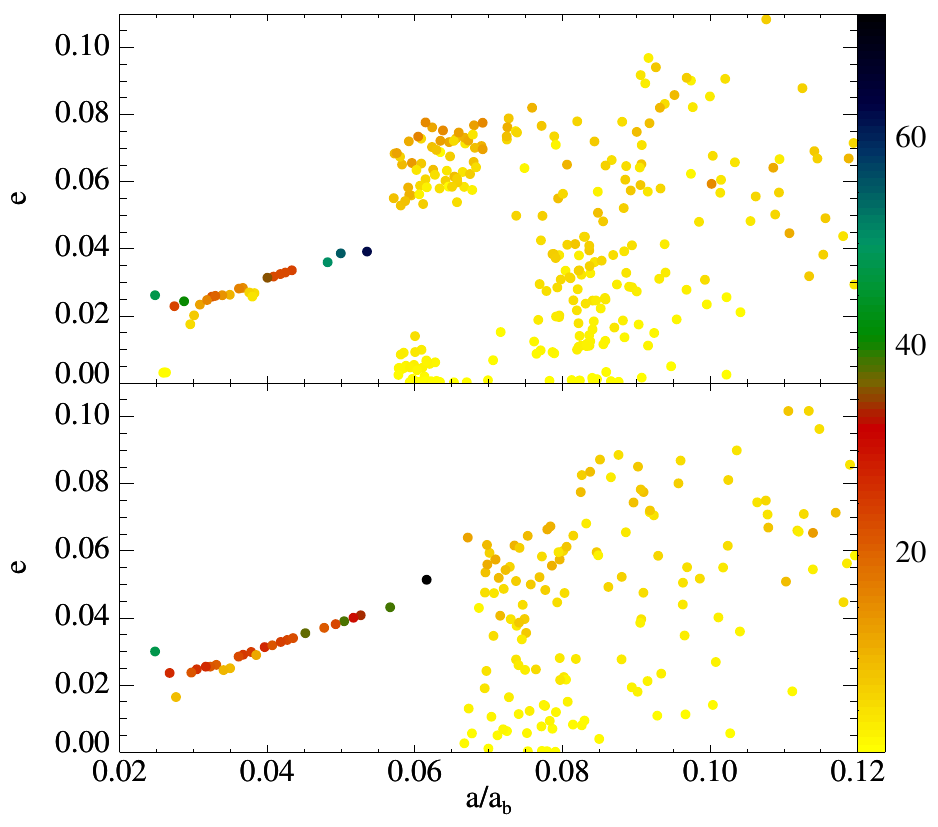}} 
\caption{Eccentricity vs. semi-major axis of test particles in the $\alphacen$ system after 1000 binary orbits. Color indicates size in km. Top panel: $\alphacen$ A. Bottom panel: $\alphacen$ B.} 
\label{figaealphacen} 
\end{figure} 

\begin{figure} 
\resizebox{\hsize}{!}{\includegraphics{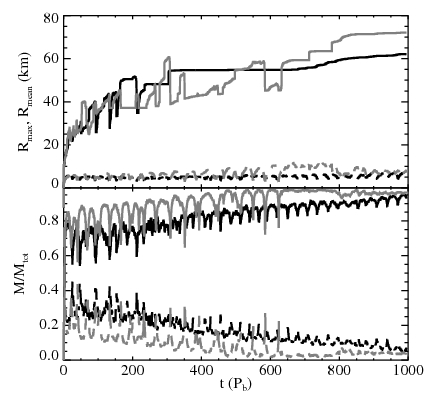}} 
\caption{Evolution of maximum particle size (solid curve, top panel), mean particle size (dashed curve, top panel), planetesimal mass fraction (solid curve, bottom panel) and dust mass fraction (dashed curve, bottom panel) for the $\alphacen$ system. Black curves are for $\alphacen$ A, grey curves are for $\alphacen$ B.} 
\label{figsizemassalphacen} 
\end{figure} 

We consider the $\alphacen$ system in Figs. \ref{figaealphacen} and \ref{figsizemassalphacen}. Results are shown for both binary components A and B, but since only the binary mass ratio changes between the two, the results are very similar. Again, we see growth up to $70$ km in the inner region of the disc. Due to the fact that the system is more strongly perturbed than $\gamceph$, the accretion-friendly region has shifted inward compared to Fig. \ref{figae}. The rough periodicity seen at early times in the bottom panel of Fig. \ref{figsizemassalphacen} is due to different generations of planetesimals. The largest objects are formed inside $1.4$ AU with significant growth inside habitable zone $< 1$ AU. 

\section{Discussion}
We present the first study of planet formation in binaries taking into account the physical effects of collisions.  We show that these effects tend to favour growth towards large planetesimals in the perturbed system. Two main mechanisms have been identified. First, frequent collisions tend to prevent planetesimals from reaching their equilibrium orbits. If collision rates are high enough, this means that the accretion-hostile environment due to differential orbital phasing is never reached. Second, fragments produced by high-velocity collisions make up a large reservoir of material that is very easily reaccreted by the remaining planetesimals as they sweep through the disc on highly eccentric orbits. 
 
We have chosen parameters that in some respects should be unfavourable for planetesimal growth. Gas drag is strong throughout the disc in the models presented here. However, since gas drag plays only a minor role, changing the gas density to a more realistic power law in radius does not change the results. New planetesimals are formed on circular orbits and are, therefore, immediately capable of destroying larger bodies that are on eccentric orbits. Forming new planetesimals on eccentric equilibrium orbits would favour more accreting collisions. On the other hand, we have assumed dust accretion to be efficient. This efficiency depends on the size distribution that is produced in collisions and the vertical extent over which the fragments are distributed. Some of the fragments may be lost due to radial drift, which is ignored in the current model, but it may also help bringing more mass into the accretion friendly inner regions. 

We have worked in a 2D geometry for computational reasons. Although we have tried to scale the surface density in such a way to get realistic collision time scales, it is important to realise that $\tauctd$ evolves in a different way with particle size than $\tauc$. It is, therefore, not possible to have realistic collision time scales at all times. Furthermore, inclinations of the planetesimals may be excited by collisions, resulting in $i=e/2$, which decreases the collision time scale. However, since most collisions are destructive, it is likely that the remaining planetesimals will essentially orbit in the plane of the disc. Other important three-dimensional effects were pointed out by \citet{xie09} and \citet{xie10}, who showed that if the disc is inclined with respect to the binary plane, this may favour planetesimal accretion. Clearly, three-dimensional simulations are the way forward, and the results in this letter should be interpreted as a proof of principle, that planetesimal accretion might be possible in tight binary systems under the right circumstances.

\section*{Acknowledgements}
We thank Philippe Th\'ebault for useful comments that greatly improved the manuscript, Debra Fischer for getting us interested (again), and the Isaac Newton Institute for their hospitality. SJP and ZML are supported by STFC Postdoctoral Fellowships.

\bibliography{collision}

\label{lastpage}

\end{document}